\documentclass
[superscriptaddress,secnumarabic,amssymb,amsmath,nobibnotes,aps,prd,showkeys,nofootinbib,nopacsnumber,onecolumn,10pt]{revtex4}%
\usepackage{graphicx}
\usepackage{bm}
\usepackage{amsmath}
\usepackage{amsfonts}
\usepackage{amssymb}%
\usepackage[utf8]{inputenc}
\providecommand{\U}[1]{\protect\rule{.1in}{.1in}}
\newcommand{\be}{\begin{equation}}
\newcommand{\ee}{\end{equation}}

\newcommand{\mincir}{\raise
-3.truept\hbox{\rlap{\hbox{$\sim$}}\raise4.truept\hbox{$<$}\ }}
\newcommand{\magcir}{\raise
-3.truept\hbox{\rlap{\hbox{$\sim$}}\raise4.truept\hbox{$>$}\ }}

\begin{document}

\title{Anisotropic Spacetimes in $f(G)$-gravity: Bianchi I, Bianchi III and Kantowski-Sachs Cosmologies}

\author{R. S. Bogadi}
\email{robertb@dut.ac.za}
\affiliation{Department of Mathematics, Faculty of Applied Sciences, Durban University of Technology, Durban 4000, South Africa}

\author{A. Giacomini}
\email{alexgiacomini@uach.cl}
\affiliation{Instituto de Ciencias Fisicas y Matem\`{a}ticas, Universidad Austral de Chile, Valdivia, Chile}

\author{M. Govender}
\email{megandhreng@dut.ac.za}
\affiliation{Department of Mathematics, Faculty of Applied Sciences, Durban University of Technology, Durban 4000, South Africa}

\author{C. Hansraj}
\email{hansrajc@sun.ac.za}
\affiliation{Applied Mathematics Division, Department of Mathematical Sciences, 
Stellenbosch University, Private Bag X1, Matieland 7602, South Africa}

\author{G. Leon}
\email{genly.leon@ucn.cl}
\affiliation{Departamento de Matem\`{a}ticas, Universidad Cat\`{o}lica del Norte, Avda. Angamos 0610, Casilla 1280 Antofagasta, Chile}
\affiliation{Department of Mathematics, Faculty of Applied Sciences, Durban University of Technology, Durban 4000, South Africa}

\author{A. Paliathanasis}
\email{anpaliat@phys.uoa.gr}
\affiliation{Department of Mathematics, Faculty of Applied Sciences, Durban University of Technology, Durban 4000, South Africa}
\affiliation{Departamento de Matem\`{a}ticas, Universidad Cat\`{o}lica del Norte, Avda. Angamos 0610, Casilla 1280 Antofagasta, Chile}
\affiliation{National Institute for Theoretical and Computational Sciences (NITheCS), South Africa.}

\begin{abstract}

We investigate the evolution of cosmological anisotropies within the framework of $f\left(G\right)$-gravity. Specifically, we consider a locally rotationally symmetric geometry in four-dimensional spacetime that describes the Bianchi I, Bianchi III, and the Kantowski-Sachs spacetimes. Within this context, we introduce a Lagrange multiplier which allows us to reformulate the geometric degrees of freedom in terms of a scalar field. The resulting theory is dynamically equivalent to an Einstein-Gauss-Bonnet scalar field model. We normalize the field equations by introducing dimensionless variables. The dynamics of our system is then explored by solving the resulting nonlinear differential equations numerically for various sets of initial conditions. Our analysis reveals the existence of two finite attractors: the Minkowski universe and an isotropic, spatially flat solution capable of describing accelerated expansion. Although de Sitter expansion may be recovered, it appears only as an unstable solution. In addition, the theory suffers from the existence of Big Rip singularities.

\end{abstract}


\date{\today}

\maketitle

\section{Introduction}

The generalization of General Relativity (GR) to higher-dimensional geometries is described according to Lovelock's theory \cite{lov1}, in which new scalars related to the Riemann tensor are introduced into the Lagrangian of the action integral. Lovelock's theory of gravity reduces to GR when the background geometry is four-dimensional and, in general, constitutes a second-order theory of gravity, meaning that the theory is free from Ostrogradsky instabilities. Modified gravity can then be considered to be the more generalized $f(Lovelock)$-gravity \cite{bue} as the various scalars arise when the Lagrangian is extended to higher order in the Euler densities. This is actively studied in cosmology \cite{noj1}. The Gauss-Bonnet scalar is the scalar which follows after the Ricci scalar in the extended Euler densities as given in Lovelock's theory, and this defines Gauss-Bonnet gravity \cite{pan1}. The Gauss-Bonnet term introduces new geometrodynamical degrees of freedom into the field equations, allowing for additional properties of physical space to be generated \cite{pp1,pp2,pp3,pp4,pp5,pp6,pp7,pp8,pp9,pp10,pp11,pp12}. Indeed it has shown recent success in describing the observed acceleration of the universe \cite{loh}. 

Gauss-Bonnet gravity in higher dimensions is actively studied in the context of cosmology because of its consistency in describing the expanding universe, its applications to astrophysics and its relation to string theory \cite{sin,dey,bog,chev0}. In the case of a four-dimensional manifold, the Gauss-Bonnet scalar is a topological invariant. However, when a scalar field is introduced and coupled to the Gauss-Bonnet scalar \cite{gg3,gg6,gg7,gg8,gg12,gg10,gg11}, or when a nonlinear function of the Gauss-Bonnet scalar \cite{Li:2007jm,g11,g12,g13,g14} is included in the Action Integral, we obtain four-dimensional theories in which the Gauss-Bonnet term contributes nontrivially. These theories have a wide range of applications in cosmology as they offer a geometric framework for describing both inflation and the late-time acceleration of the universe commonly attributed to dark energy. An alternative approach to introduce the Gauss-Bonnet scalar in the four-dimensional  geometry, is the introduction of the a rescaled coupling constant to the Gauss-Bonnet scalar by a factor of $\alpha\slash(D-4)$. Such that taking the limit $D \to 4$ allows for nontrivial higher curvature effects on the 4D local gravity \cite{glavanlin}, circumventing Lovelock theory implications. However, this consideration has been received criteria \cite{comg}. 

Recently, there have been new results in different stellar structures \cite{chakraborty1, banerjee1, tangphati1, hansrajs1, doneva1, hansrajs2, hansrajs3}, gravitational lensing \cite{islam1, kumar1, babar1, islam2} and a series of black hole configurations \cite{ghosh1, konoplya1, ghosh2, konoplya2, dveer1, dveer2, wei1, yang1, fernandes1, zhang1, mansoori1, ali1, ladghami1} including shadows \cite{konoplya3, guo1, zeng1}. Quasinormal modes and stability \cite{churilova, mishra1, aragon1, jafarzade}, as well as wormholes \cite{jusufi1, liu1, panyasiripan, godani1} were also explored. For a detailed review of 4D EGB theory and its physical implications in the cases of black holes, cosmology and strong/weak gravitational fields, the reader is referred to \cite{fernandes2}.

Papagiannopoulos {\it et al.} \cite{papa} investigated in detail the phase space of the gravitational field equations within the isotropic and homogeneous Friedmann--Lema\^{\i}tre--Robertson--Walker (FLRW) cosmology using $f\left( G\right)$-gravity theory \cite{Li:2007jm}. It was found that the theory can describe de Sitter expansion, but it is unlikely
that the full evolution of the universe could be accounted for as it suffers from a fine-tuning problem. In this work, we focus on analyzing the evolution of anisotropies. Specifically, we consider the background geometry to be described by a locally rotationally symmetric (LRS) spacetime of the Bianchi I, Bianchi III, or Kantowski-Sachs family.

Anisotropies play an important role in the early stages of the universe \cite{ani1,ani2}. Thus through this study, we examine whether $f\left(  G\right) $-gravity can offer a geometric solution to the isotropization of the universe, as well as elucidating the flatness problem. On the other hand, it was found in \cite{antop} that within Einstein-Gauss-Bonnet scalar field theory (without a kinetic term), the evolution of anisotropies leads to the compactification of spacetime. Therefore, it is of particular interest to investigate whether similar behavior arises in the case of $f\left( G\right) $-gravity.

In Section \ref{sec2} we introduce the gravitational framework for our cosmological model, namely $f\left( G\right)$-gravity in four-dimensional spacetime. We introduce a Lagrange multiplier and then show that this gravitational model is equivalent to Einstein-Gauss-Bonnet scalar field theory without the scalar field kinetic term. For the background geometry we consider an LRS homogeneous geometry which describes the Bianchi I, the Bianchi III and the Kantowski-Sachs spacetimes. We consider Misner-like variables and in Section \ref{sec3} we show that the theory has a minisuperspace description. The cosmological field equations for an anisotropic universe are deduced with the inclusion of dimensionless variables to aid the solution process. Numerical methods are used to provide solutions for exploring the structure of the phase space. We find that the isotropic
and spatially flat universe is supported by the theory and is independent of the initial condition for the spatial curvature. It is important to note the absence of Milne-like isotropic solutions. Lastly we finalize our conclusions in Section \ref{sec4}.

\section{$f\left(G\right)$-gravity} \label{sec2}

We consider a four-dimensional Riemannian manifold with metric tensor $g_{\mu\nu}$ and covariant derivative $\nabla_{\mu}$ defined by the Levi-Civita connection $\Gamma_{\mu\nu}^{\kappa}$. The curvature tensor, Ricci tensor and Ricci scalar are defined as%

\begin{align}
R_{~\lambda\mu\nu}^{\kappa}  & = 2\partial_{\lbrack\mu}\Gamma_{\;\;\nu]\lambda}^{\kappa}+2\Gamma_{\;\;[\mu|\sigma|}^{\kappa}\Gamma_{\;\;\nu]\kappa}^{\sigma}~, \\
R_{\mu\nu}  & = 2\partial_{\lbrack\kappa}\Gamma_{\;\;\mu]\nu}^{\kappa} + 2\Gamma_{\;\;[\kappa|\sigma|}^{\kappa}\Gamma_{\;\;\mu]\nu}^{\sigma}~, \\
R  & = 2g^{\mu\nu}\partial_{\lbrack\kappa}\Gamma_{\;\;\mu]\nu}^{\kappa} + 2g^{\mu\nu}\Gamma_{\;\;[\kappa|\sigma|}^{\kappa}\Gamma_{\;\;\mu]\nu}^{\sigma}~,
\end{align}

from which the Gauss-Bonnet scalar is given by%

\begin{equation}
G = R^{2} - 4R_{\mu\nu}R^{\mu\nu} + R_{\mu\nu\sigma\lambda}R^{\mu\nu\sigma\lambda}.
\end{equation}

This scalar is a topological invariant on a four-dimensional manifold. The framework for the gravitational field is then taken to be the modified Gauss-Bonnet theory as determined according to the action integral \cite{Li:2007jm}

\begin{equation}
S_{f\left(  G\right) } = \int d^{4}x\sqrt{-g}\left( R + f\left(  G\right)\right) . \label{cc.02}%
\end{equation}

This integral modifies the Einstein-Hilbert action by incorporating $f\left( G\right) $ as an arbitrary function of the Gauss-Bonnet scalar. In the case where $f\left( G\right) $ is a linear function, the limit of General Relativity is recovered. \\
In the case of vacuum, variation with respect to the metric tensor leads to the modified gravitational field equations \cite{Li:2007jm},

\begin{align}
R_{\mu\nu}-\frac{R}{2}g_{\mu\nu} & = \frac{1}{2}g_{\mu\nu}f - 2f,_{G}RR_{\mu\nu} + 4f_{,G}R_{\mu}^{~~\lambda}R_{\nu\lambda} - 2f_{,G}R_{\mu\kappa\lambda\sigma}R_{\nu}^{~\kappa\lambda\sigma} - 4f_{,G}R_{\mu\kappa\lambda\nu}R^{\kappa\lambda}  \nonumber \\
&  + 2\left[  R\left(  \delta_{\mu}^{\kappa}\delta_{\nu}^{\lambda}-2g_{\mu\nu}g^{\kappa\lambda}\right)  + 2\left(  R_{\mu\nu}g^{\kappa\lambda} + g_{\mu\nu}R^{\kappa\lambda}\right)  \right] \nabla_{\kappa}\nabla_{\lambda}f_{,G} \nonumber \\
&  - 8R_{(\mu|\kappa}\nabla_{\nu)}\nabla^{\kappa}f_{,G} - 4R_{\mu\lambda\nu\sigma}\nabla^{\kappa}\nabla^{\lambda}f_{,G}
\end{align}

where the left hand side is the Einstein tensor and the right hand side describes the geometrodynamical degrees of freedom introduced by the nonlinear $f\left(G\right)$ function.

\subsection{Equivalency with Einstein-Gauss-Bonnet scalar field theory}

Scalar fields have been widely used in modified theories of gravity to describe the geometrodynamical degrees of freedom. In a similar approach, by introducing a Lagrange multiplier $\lambda$ within the action integral (\ref{cc.02}) we can write the field equations in the equivalent form of a scalar field theory.

Consider the constraint equation $G=\mathcal{G}$, where $\mathcal{G}$ is the functional form of the Gauss-Bonnet scalar. Then the action integral (\ref{cc.02}) can be written in the equivalent form \cite{papa},

\begin{equation}
S_{f\left(  G\right)  }=\int d^{4}x\sqrt{-g}\left(  R+f\left(  G\right) - \lambda\left(  G-\mathcal{G}\right)  \right) . \label{cc.03}%
\end{equation}

Variation with respect to the Gauss-Bonnet scalar gives the constraint $\lambda=f_{,G}\left(  G\right)$. Therefore, the action (\ref{cc.03}) is expressed as

\begin{equation}
S_{f\left(  G\right)  }=\int d^{4}x\sqrt{-g}\left(  R+ \left(  f\left(G\right)  - G f_{,G}\left( G\right)  \right)  + f_{,G}\left(  G\right)\mathcal{G}\right) ,
\end{equation}

or in the equivalent form \cite{papa},

\begin{equation}
S_{f\left(  G\right) } = \int d^{4}x\sqrt{-g}\left(  \frac{R}{2}+\phi G + V\left(  \phi\right) \right)~, \label{cc.04}%
\end{equation}

where we have introduced the scalar field $\phi$ and the potential function $V\left( \phi\right)$. This is defined as,

\begin{subequations}
\begin{align}
&  \phi = f_{,G}\left(  G\right)  ,\label{cc.05}\\
&  V\left(  \phi\right)  = \left(  f\left(  G\right)  -Gf_{,G}\left(  G\right)\right) . \label{cc.06}%
\end{align}

The gravitational model (\ref{cc.04}) belongs to the family of Einstein-Gauss-Bonnet scalar field theory \cite{ss1,ss2,ss3,ss4} where the kinetic term has been omitted. This is similar to the equivalency of the $f\left( R\right) $-theory with the Brans-Dicke theory in which a zero Brans-Dicke parameter results in its omittance. Furthermore, from expressions (\ref{cc.05}), (\ref{cc.06}) it follows that

\end{subequations}
\begin{equation}
f\left(  G\right) = V\left(  \phi\left(  f\left(  G\right)  \right)  \right) - \phi\left(  f\left(  G\right)  \right)  V_{,\phi}\left(  \phi\left(  f\left(G\right)  \right)  \right) . \label{cc.07}%
\end{equation}

\section{Anisotropic cosmologies}  \label{sec3}

In this work we consider a background geometry that is described by an anisotropic and homogeneous line element expressed in terms of Misner variables \cite{book1}, 

\begin{equation}
ds^{2} = -N^{2}\left( t\right) dt^{2} + e^{2a}\left( e^{2\beta\left( t\right)}dx^{2} + e^{-\beta\left(  t\right)  }\left(  dy^{2} + S^{2}\left(  y;\kappa\right) dz^{2}\right) \right) , \label{lrs.01}%
\end{equation}

in which $\kappa=0,\pm1$, and $S\left(  y;0\right) = 1,~S\left(  y;1\right)  =\sinh y,~S\left(  y;-1\right) = \sin y $. Parameter $\beta\left(  t\right)$ describes anisotropy and $N\left(  t\right) $ is the lapse function. The volume is given by $V\left( t\right)  =  N\left(  t\right) e^{3a\left(  t\right)  }$. For the comoving observer, $u^{\mu}=\frac{1}{N}\delta_{t}^{\mu}$ and the expansion rate is defined as $\theta = 3H = \frac{1}{N}\dot{a}$, where $H$ is the average Hubble function. \\

We observe that for $\kappa=0$, the line element (\ref{lrs.01}) admits a four-dimensional Lie algebra, which form the $A_{1}\otimes A_{3,4}$ in the Patera {\it et al.} classification scheme \cite{patera}. Thus, the line element (\ref{lrs.01}) describes the LRS Bianchi I geometry \cite{book1}. For $\kappa=1$, the line element (\ref{lrs.01}) admits the four-dimensional Lie algebra, which form the $A_{1}\otimes A_{3,5}$ \cite{patera}, and this gives us the Bianchi III geometry \cite{book1}.
Lastly, for $\kappa=-1$, the admitted four isometries of the spacetime (\ref{lrs.01}) form the $A_{4,10}$ \cite{patera} so that the line element in this case describes the Kantowski-Sachs geometry \cite{book1}.

For the line element (\ref{lrs.01}) the following expressions for the Ricci
and the Gauss-Bonnet scalars are obtained,

\begin{equation}
R = 6\left( \frac{1}{N}\dot{H}+2H^{2}\right)  +\frac{3}{2N^{2}}\dot{\beta}^{2}-\kappa e^{\beta-2a}, \label{rr.01}%
\end{equation}%

\begin{align}
G & =\frac{6}{N^{5}}\left(  \dot{\beta}-2HN\right)  \left(  \left( HN-\dot{N}\right)  \dot{\beta}^{2}+N\dot{\beta}\left( \ddot{\beta}-\dot{H}N-H^{2}N^{2}\right)  -2HN^{3}\left(  H^{2}N+\dot{H}\right) \right) \nonumber\\
& -8\kappa e^{\beta-2a}\left(  H^{2}+\frac{1}{N^{2}}\left(  \dot{\beta}^{2}+\ddot{\beta}-\frac{\dot{N}}{N^{2}}\dot{\beta}\right)  +\frac{1}{N}\left(\dot{H}+2H\dot{\beta}\right)  \right) . \label{rr.02}%
\end{align}

By replacing expressions (\ref{rr.01}) and (\ref{rr.02}) in the action integral (\ref{cc.04}) and integrating by parts, we end up with the point-like Lagrangian for the field equations,

\begin{equation}
L\left( N,a,\dot{a},\beta,\dot{\beta},\phi,\dot{\phi}\right) = e^{3a}\left(-\frac{3}{N}\dot{a}^{2} + \frac{3}{4}\dot{\beta}^{2} + \frac{4}{N^{3}}\dot{a}^{3}\dot{\phi} - \frac{3}{N^{3}}\dot{a}\dot{\beta}^{2}\dot{\phi} + \dot{\beta}^{3}\dot{\phi} - NV\left(\phi\right) \right) + \kappa e^{\beta+a}.
\label{lrs.02}%
\end{equation}

Indeed, when anisotropy is eliminated, $\dot{\beta}=0$, the point-like Lagrangian of FLRW spacetime is recovered \cite{papa}.

The cosmological field equations follow from the variation of the Lagrangian function (\ref{lrs.02}) with respect to the dynamical variables $\left\{N,a,\beta,\phi\right\} $. It is important to mention that variation with respect to the lapse function, $N$, provides the constraint equation, while variation with the scalar field gives the definition for the Gauss-Bonnet
scalar $G$.

Without loss of generality, we assume the lapse function to be a constant, i.e. $N\left( t\right) = 1$, such that the field equations read

\begin{equation}
0=3H^{2}-\frac{3}{4}\dot{\beta}^{2}-12H^{3}\dot{\phi}+9H\dot{\beta}^{2}%
\dot{\phi}-3\dot{\beta}^{3}\dot{\phi}-V\left(  \phi\right)  -\kappa
e^{\beta-2a}, \label{lrs.03}%
\end{equation}%

\begin{align}
0  &  =2\dot{H}-8H^{3}\dot{\phi}+\dot{\beta}^{3}\dot{\phi}-8H\dot{H}\dot{\phi
}+2\dot{\beta}\dot{\phi}\ddot{\beta}\nonumber\\
&  +H^{2}\left(  3-4\ddot{\phi}\right)  +\dot{\beta}^{2}\left(  \frac{3}%
{4}+\ddot{\phi}\right)  +\frac{\kappa}{3}e^{\beta-2\alpha}-V\left(
\phi\right)  ,
\end{align}%

\[
0=-\frac{3}{2}\ddot{\beta}+18H^{2}\dot{\beta}\dot{\phi}+6\dot{\beta}\dot{\phi
}\left(  \dot{H}-\ddot{\beta}\right)  -3\dot{\beta}^{2}\ddot{\phi}+\frac{3}%
{2}\left(  4\dot{\phi}\ddot{\beta}-6\dot{\beta}^{2}\dot{\phi}+\dot{\beta
}\left(  4\ddot{\phi}-3\right)  \right)  ,
\]%

\begin{equation}
0=V_{,\phi}+3\left(  2H-\dot{\beta}\right)  \left(  2H^{3}+\dot{H}\dot{\beta
}-H\left(  \dot{\beta}^{2}-2\dot{H}\right)  +\dot{\beta}\left(  \dot{H}%
-\ddot{\beta}\right)  \right)  .
\end{equation}

We introduce the dimensionless variables \cite{papa}%

\begin{align}
x  &  =\frac{\dot{\phi}}{\sqrt{1+H^{2}}}~,~y=\frac{V\left(  \phi\right)
}{1+H^{2}}~,~\Sigma=\frac{\dot{\beta}}{\sqrt{1+H^{2}}}~,~\\
\eta &  =\frac{H}{\sqrt{1+H^{2}}},~\Omega_{\kappa}=\frac{\kappa}{3}%
\frac{e^{\beta-2a}}{1+H^{2}},~\lambda=\frac{V_{,\phi}\left(  \phi\right)
}{V\left(  \phi\right)  },
\end{align}

and the new independent variable,

\begin{equation}
d\tau=\sqrt{1+H^{2}}dt~.
\end{equation}

In terms of these variables, the field equations are expressed as a set of first-order algebraic differential equations which allows for the investigation of solutions where $H\rightarrow0$, that is, a static universe.

These dimensionless variables have been widely used for the analytic study of the phase-space in cosmology. However, due to the nonlinearity of the field equations in this study, we investigate numerical simulations of the field equations for various values of the initial conditions. This allows us to study and understand the general behaviour of the model, and search for the existence of stationary points within the phase-space. For an arbitrary value of $\kappa$, the dynamical variables are six in total however due to the constraint equation, the dimension of the dynamical system can be reduced by one. In addition, for the exponential potential $V\left(  \phi\right) = V_{0}e^{\lambda\phi}$, parameter $\lambda$ is always a constant and so the dimension of the dynamical system is four.

We now proceed in presenting our numerical solutions for the field equations for the three values of parameter $\kappa$, that is, $\kappa=0,\pm 1$. \ We study the existence of asymptotic solutions for the phase-space by exploring the parametric space of the initial conditions.

\subsection{Bianchi I\qquad\qquad}

For the Bianchi I geometry, we consider $\kappa=0$, and for the exponential potential, we solve numerically the dynamical system with respect to the
variables $\left\{  x,\Sigma,\eta\right\}  $, where $\left\{  x_{0},\Sigma_{0},\eta_{0}\right\}  $ are the initial conditions. For parameter $\lambda$ we consider $\lambda<0$, in particular $\lambda=-0.1$.

\begin{figure}[ptbh]
\centering\includegraphics[width=0.8\textwidth]{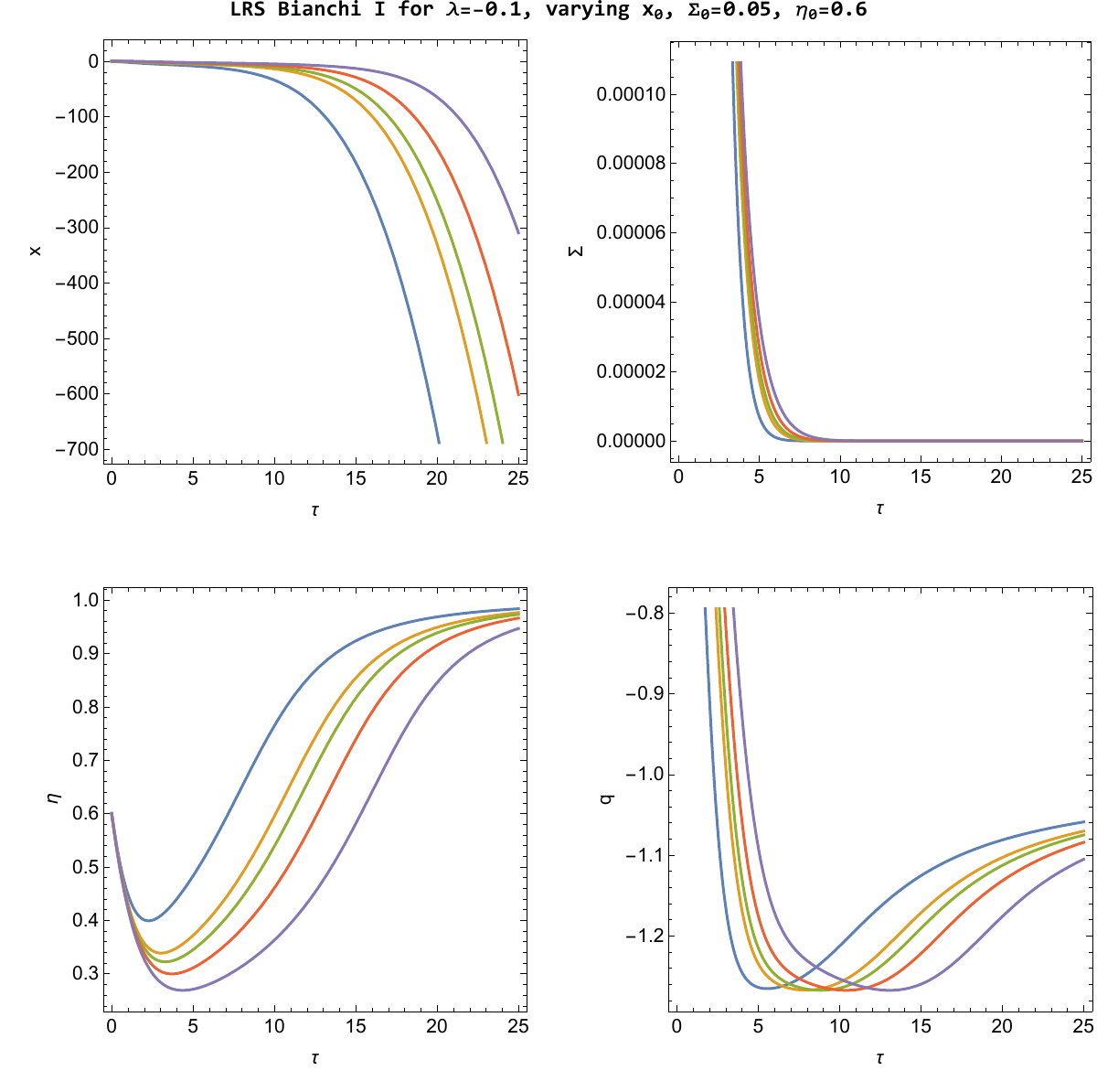}\caption{Bianchi I:
Numerical solution of the field equations within the Bianchi I background
($\kappa=0$) for different set of initial conditions. For the numerical
simulations we selected $\lambda=-0.1$. For the initial conditions we consider
$\Sigma_{0}=0.05$, $\eta_{0}=0.6$ and $x_{0}=\left\{
-0.5,0.01,0.1,0.2,0.3\right\}  $. }%
\label{b1p1}%
\end{figure}

In Fig \ref{b1p1} we plot the numerical solution of the dynamical variables for anisotropic initial conditions with different values of $x_{0}=\left\{-0.5,0.01,0.1,0.2,0.3\right\}  $, $\Sigma_{0}=0.05$ and $\eta_{0}=0.6$. We observe that the trajectories reach an asymptotic solution with $\eta\rightarrow1$ and deceleration parameter $q\rightarrow-1.1$. Recall that the deceleration parameter is defined as $q=-1-\frac{\dot{H}}{H^{2}}$.
Consequently, for this set of initial conditions, we can conclude that there exists an attractor which describes an accelerated, isotropic and homogeneous universe described by the spatially flat FLRW. The value of the deceleration parameter at the attractor depends on the constant parameter $\lambda$.

\begin{figure}[ptbh]
\centering\includegraphics[width=0.8\textwidth]{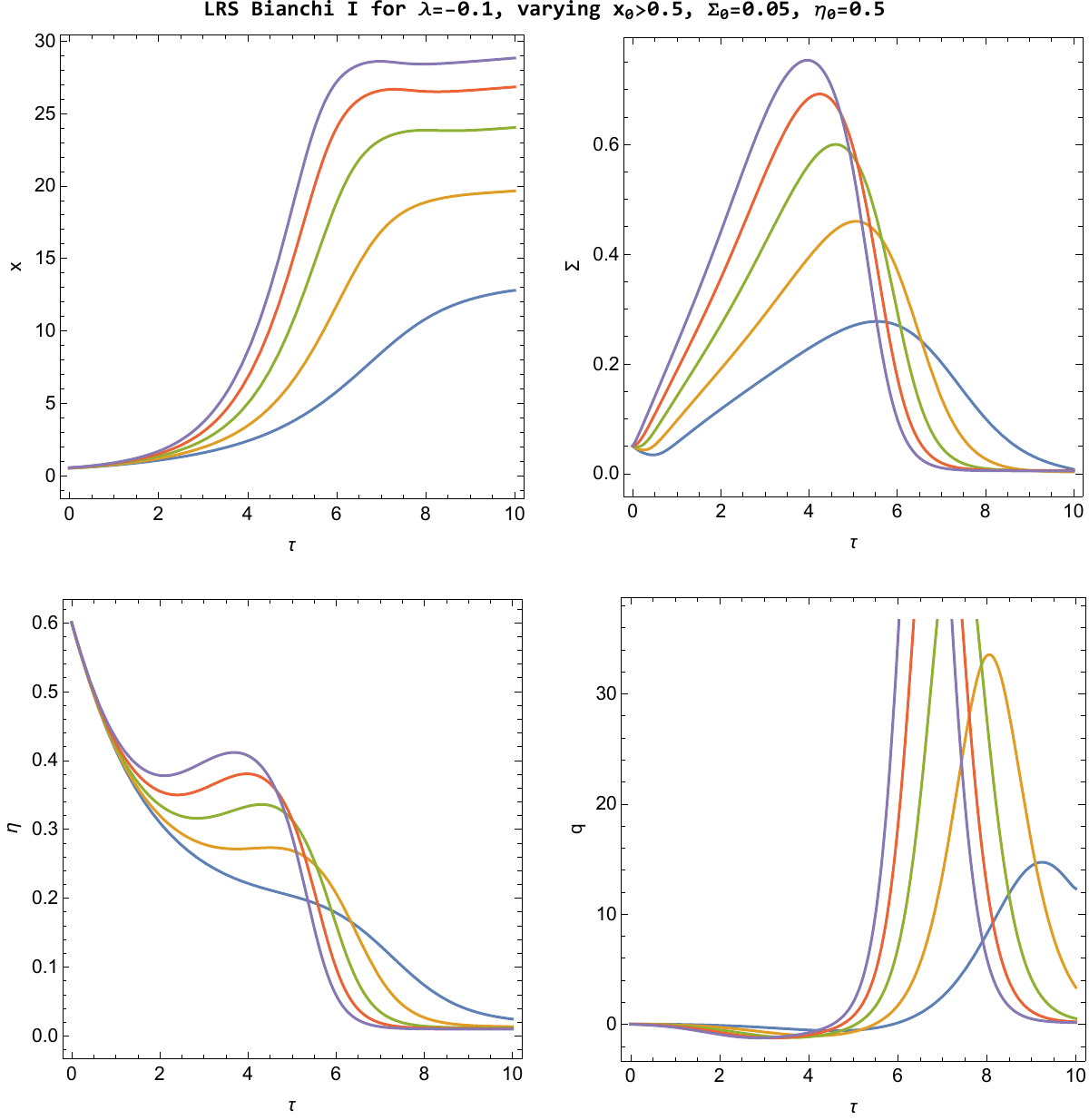}\caption{Bianchi I:
Numerical solution of the field equations within the Bianchi I background
($\kappa=0$) for different set of initial conditions. For the numerical
simulations we selected $\lambda=-0.1$. For the initial conditions we consider
$\Sigma_{0}=0.05$, $\eta_{0}=0.6$ and $x_{0}=\left\{
0.51,0.52,0.53,0.54,0.55\right\}  $. }%
\label{b1p2}%
\end{figure}

Moreover, in Fig. \ref{b1p2} we plot the numerical solution of the dynamical variables for anisotropic initial conditions with different values of $x_{0}=\left\{  0.51,0.52,0.53,0.54,0.55\right\} $, $\Sigma_{0}=0.05$ and $\eta_{0}=0.6$. The behaviour of the trajectories is different from before. The attractor is a point which describes the static isotropic static solution, that is, $\eta\rightarrow0$; $\Sigma\rightarrow0$. However around $\tau=2$, we observe that the trajectories reach a saddle point which describes an anisotropic rapid expansion. Moreover, around $\tau=4$, there is a saddle point which describes a Kasner-like solution.

\begin{figure}[ptbh]
\centering\includegraphics[width=0.8\textwidth]{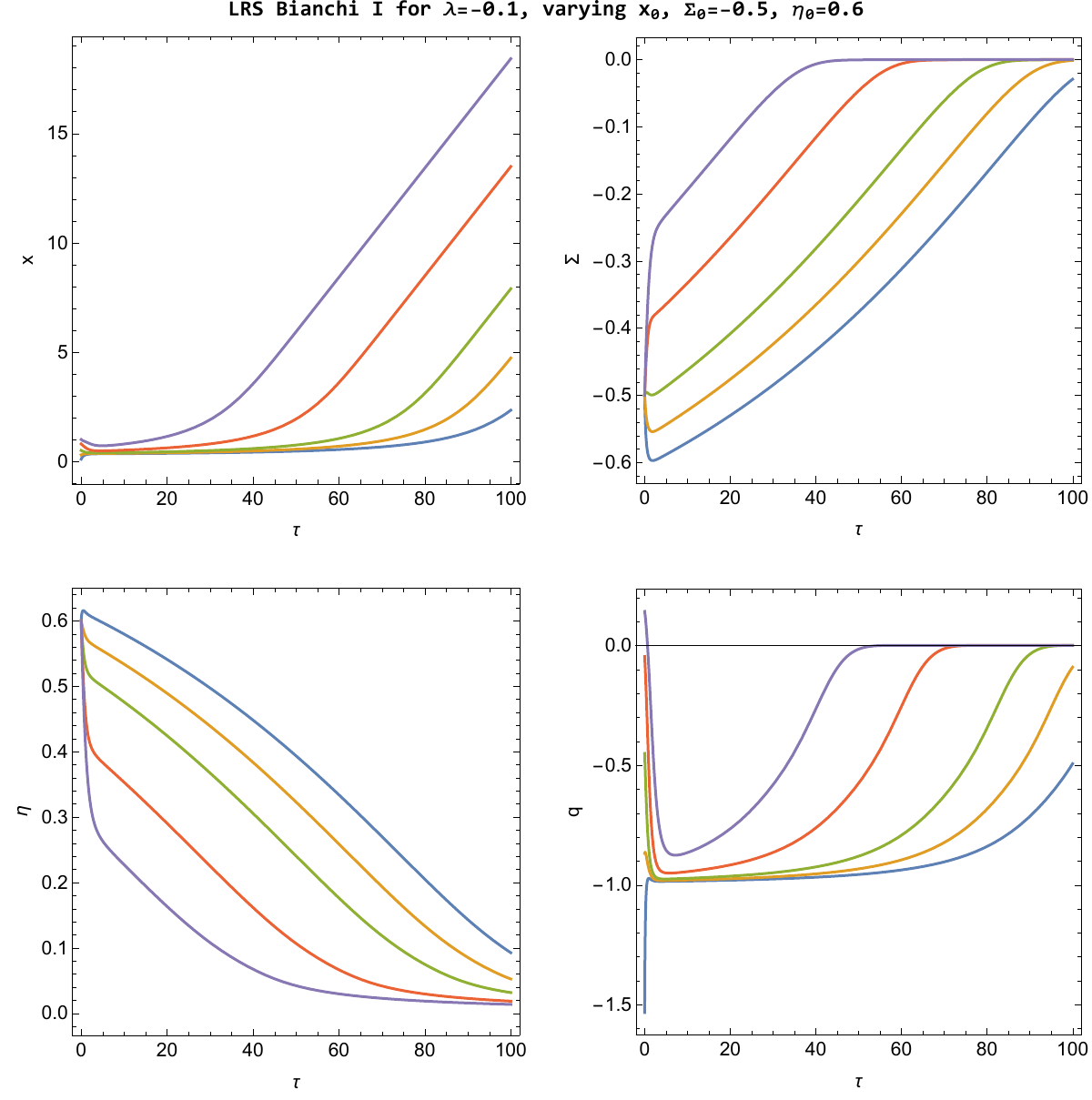}\caption{Bianchi I:
Numerical solution of the field equations within the Bianchi I background
($\kappa=0$) for different set of initial conditions. For the numerical
simulations we selected $\lambda=-0.1$. For the initial conditions we consider
$\Sigma_{0}=-0.5$, $\eta_{0}=0.6$ and $x_{0}=\left\{
0.1,0.3,0.5,0.8,1\right\}  $. }%
\label{b1p3}%
\end{figure}

In Fig. \ref{b1p3} we plot the numerical solution of the dynamical variables for anisotropic initial conditions with different values of $x_{0}=\left\{  0.1,0.3,0.5,0.8,1\right\}  $, $\Sigma_{0}=-0.5$~and $\eta_{0}=0.6$. The attractor describes the isotropic static solution, that is, $\eta\rightarrow0$; $\Sigma\rightarrow0$. The existence of an unstable asymptotic solution which describes anisotropic rapid expansion is observed in the early universe.

Finally, in Fig. \ref{b1p3} we plot the numerical solution of the dynamical variables for anisotropic initial conditions with $x_{0}=0.2$, $\Sigma_{0}=0.5$ and various values of $\eta_{0}=\left\{  0.3,0.35,0.4,0.5\right\} $. The attractor describes the isotropic and homogeneous universe as in the case of Fig. \ref{b1p1}.

\begin{figure}[ptbh]
\centering\includegraphics[width=0.8\textwidth]{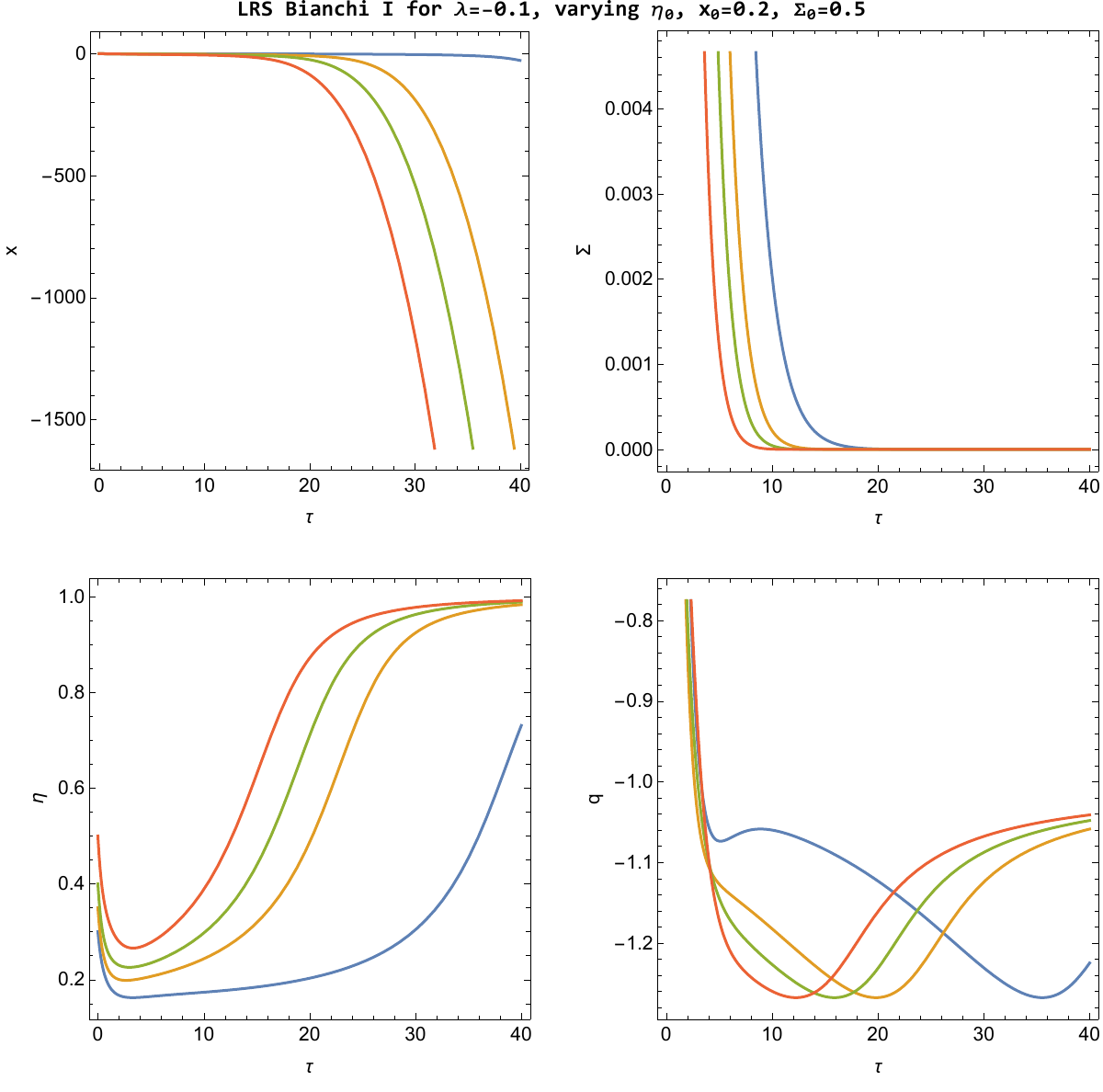}\caption{Bianchi I:
Numerical solution of the field equations within the Bianchi I background
($\kappa=0$) for different set of initial conditions. For the numerical
simulations we selected $\lambda=-0.1$. For the initial conditions we consider
$x_{0}=0.2$, $\Sigma_{0}=0.5$ and $\eta_{0}=\left\{  0.3,0.35,0.4,0.5\right\}
$}%
\label{b1p4}%
\end{figure}

From these numerical simulations we observe that for anisotropic initial conditions with zero spatial curvature, there are two attractors in the finite regime. An attractor which describes an isotropic FLRW Universe, and the anisotropic static solution. It is important to mention that for different ranges of the initial conditions, the trajectories of the dynamical variables tend to infinity which lead to Big Rip or Big Crunch singularities.

Last but not least, for $\lambda>0$, we could not find parametric space for the initial conditions $\left\{ x_{0},\Sigma_{0},\eta_{0}\right\} $ where the attractors in the finite regime exist. We determined solutions which describe Big Rip singularities.

\subsection{Bianchi III}

For $\kappa>0$, the evolution of the physical parameters is described by the evolution of the dynamical variables $\left\{  x,\Sigma,\eta,\Omega_{\kappa}\right\}$. Let us assume that $\left\{  x_{0},\Sigma_{0},\eta_{0}, \Omega_{\kappa0}\right\} $ are the initial conditions for the numerical solution of the field equations. For the constant parameter $\lambda$ we consider the value $\lambda=-0.1$.

\begin{figure}[!htbp]
\centering\includegraphics[width=0.8\textwidth]{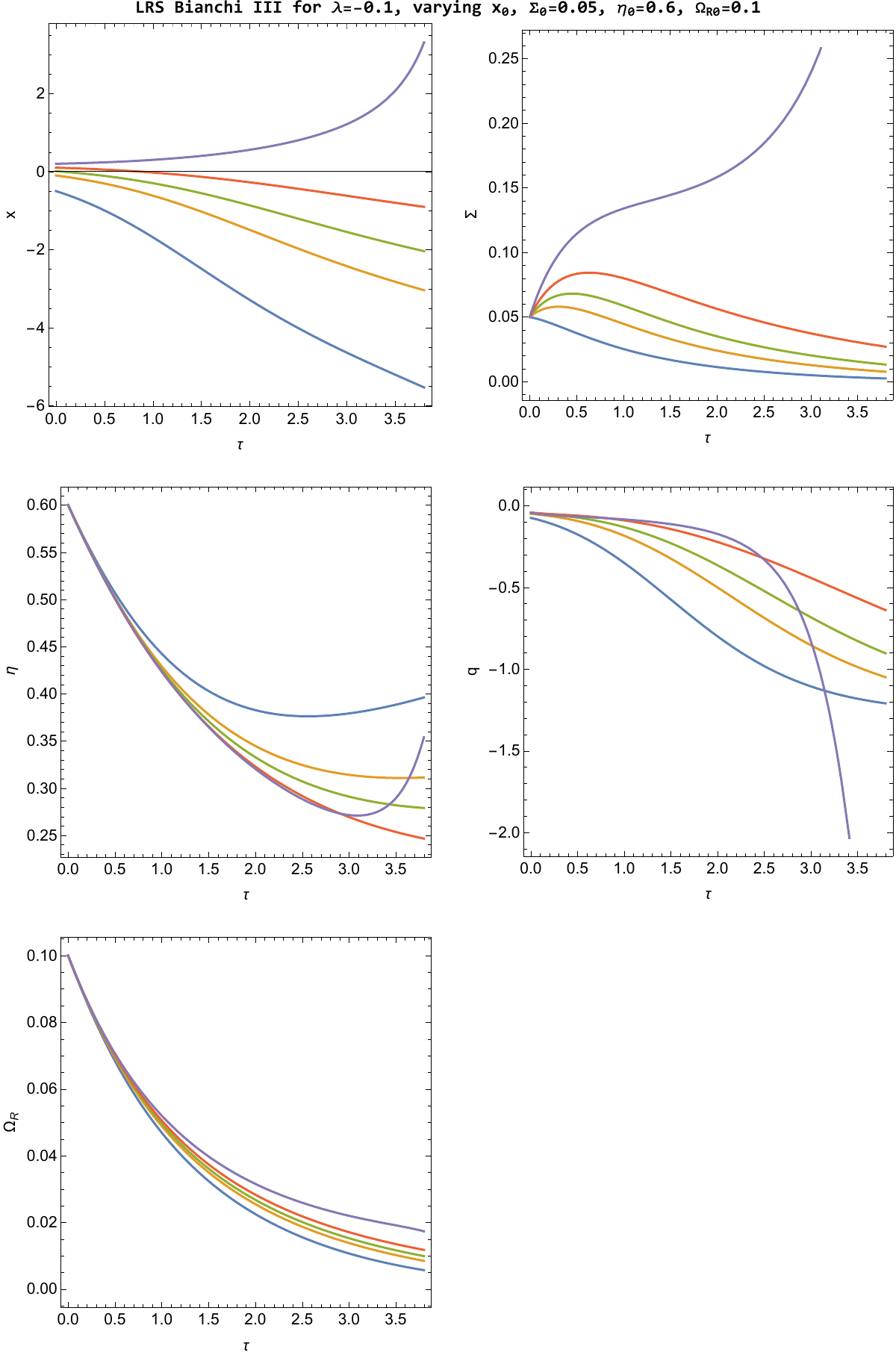}\caption{Bianchi III:
Numerical solution of the field equations within the Bianchi III background
for different set of initial conditions. For the numerical simulations we
selected $\lambda=-0.1$. For the initial conditions we consider $\Sigma
_{0}=0.05$, $\eta_{0}=0.6,\Omega_{\kappa0}=0.1$ and $x_{0}=\left\{
-0.5,-0.1,-0.01,0.1,0.2\right\}  $. }%
\label{b3p1}%
\end{figure}

In Fig. \ref{b3p1} we present numerical solutions of the field equations for the initial conditions $x_{0}=\left\{  -0.5,-0.1,-0.01,0.1,0.2\right\}$, $~\Sigma_{0}=0.05$, $\eta_{0}=0.6~$and $\Omega_{\kappa0}=0.1$. We observe that the isotropic and spatially flat FLRW universe is an attractor for the system. In particular, the asymptotic solution is that described by the solution shown in Fig. \ref{b1p1}. 

\begin{figure}[!htbp]
\centering\includegraphics[width=0.8\textwidth]{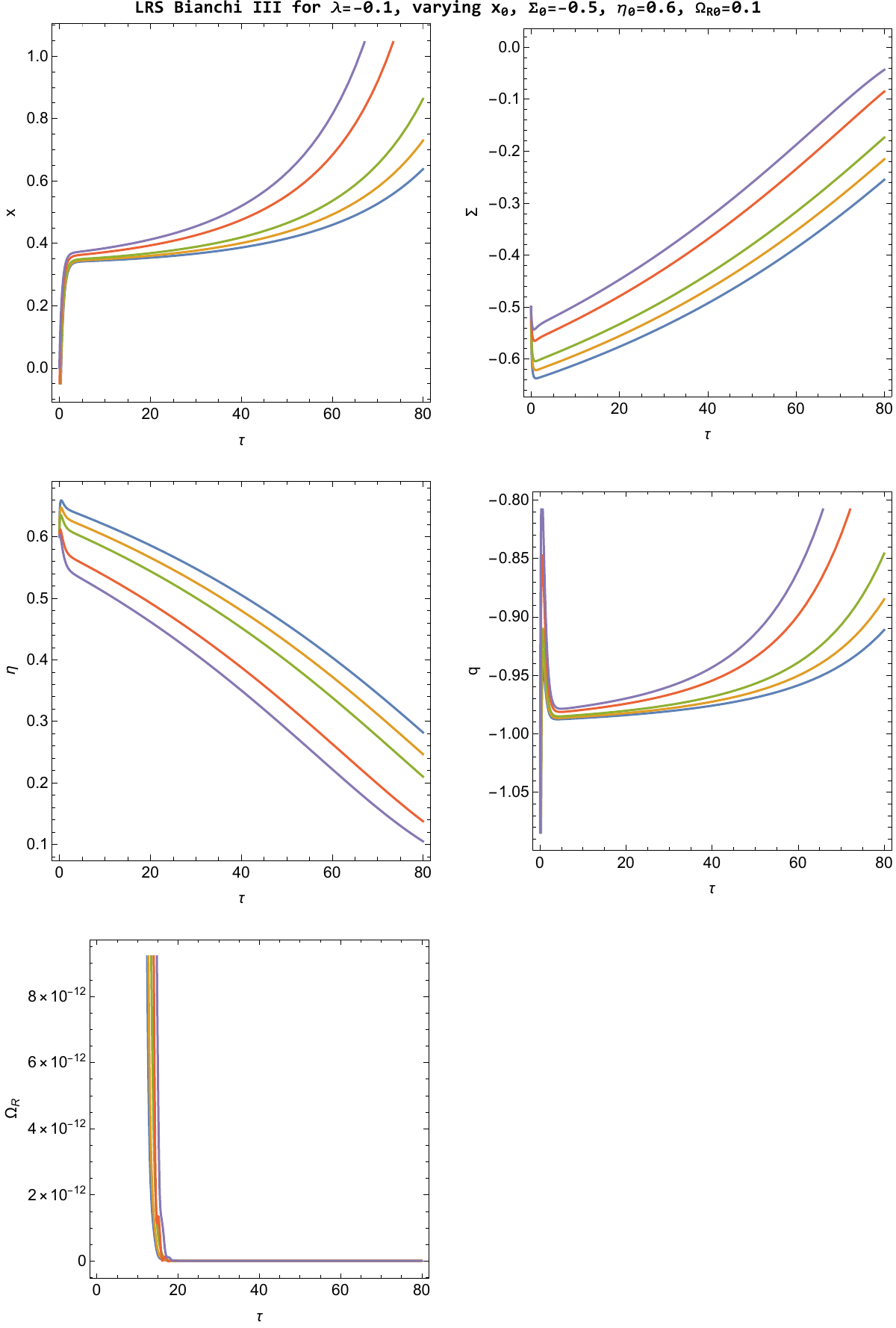}\caption{Bianchi III:
Numerical solution of the field equations within the Bianchi III background
for different set of initial conditions. For the numerical simulations we
selected $\lambda=-0.1$. For the initial conditions we consider $\Sigma
_{0}=0.05$, $\eta_{0}=0.6,\Omega_{\kappa0}=0.1$ and $x_{0}=\left\{
-0.5,-0.1,-0.01,0.1,0.2\right\}  $. }%
\label{b3p2}%
\end{figure}

Furthermore, in Fig. \ref{b3p2} we present the numerical solution for the initial conditions$~x_{0}=\left\{  -0.5,-0.4,-0.3,-0.1,0.0\right\} $, $~\Sigma_{0}=-0.5$, $\eta_{0}=0.6~$ and $\Omega_{\kappa0}=0.1.$ We observe that there exists an attractor which describes an isotropic and static universe, while an unstable solution which describes rapid expansion where the cosmological constant term dominates exists. 

With the exception of the above, Big Rip and Big Crunch singularities are determined when the dynamical parameters reach infinity. However, no other attractors were determined in the finite regime. Consequently, no solutions with nonzero spatial curvature, such as the Milne universe, were observed to be attractors.

\subsection{Kantowski-Sachs}

For initial conditions which describe a Kantowski-Sachs geometry, $\Omega_{\kappa0}<0$, \ and for $\lambda=-0.1$, we present the evolution of the dynamical variables $\left\{  x,\Sigma,\eta,\Omega_{\kappa}\right\}  $ as they follow from the numerical solution of the dynamical system for different values of the initial conditions $\left\{  x_{0},\Sigma_{0},\eta_{0} ,\Omega_{\kappa0}\right\}  $.

\begin{figure}[!htbp]
\centering\includegraphics[width=0.8\textwidth]{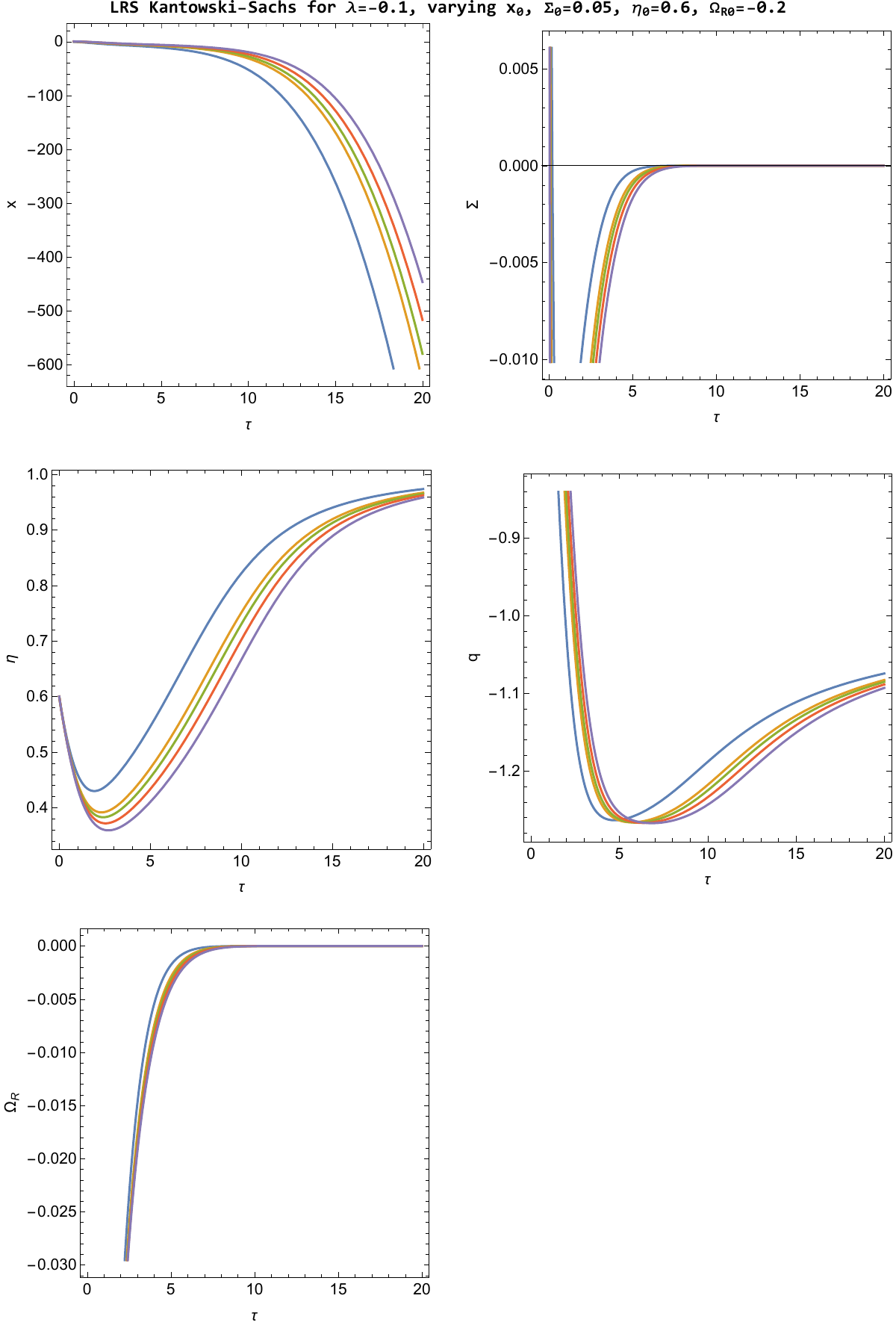}\caption{Kantowski-Sachs: Numerical solution of the field equations within the Kantowski-Sachs background for different set of initial conditions. For the numerical simulations we selected $\lambda=-0.1$. For the initial conditions we consider $\Sigma_{0}=0.05$, $\eta_{0}=0.6,\Omega_{\kappa0}=-0.2$ and $x_{0}=\left\{-0.5,0.01,0.1,0.2,0.3\right\}$. }%
\label{ksp1}%
\end{figure}

In Fig. \ref{ksp1} we consider the initial conditions $x_{0}=\left\{-0.5,0.01,0.1,0.2,0.3\right\}  $,$~\Sigma_{0}=0.05$, $\eta_{0}=0.6~$and $\Omega_{\kappa0}=-0.2$. We observe that the isotropic and spatially flat universe described by the solution in Fig. \ref{b1p1} is an attractor for the model. 

On the other hand, for the initial conditions $x_{0}=0.2,\Sigma_{0}=\left\{-1,-0.7,-0.6,-0.5.-0.4\right\}  $, $\eta_{0}=0.6$ and $\Omega_{\kappa0}=-0.2$, attractor is the static solution described by the Minkowski Universe. However, there exists a saddle point within the phase space which describes a rapid expansion as it is given by the cosmological constant term.

\begin{figure}[!htbp]
\centering\includegraphics[width=0.8\textwidth]{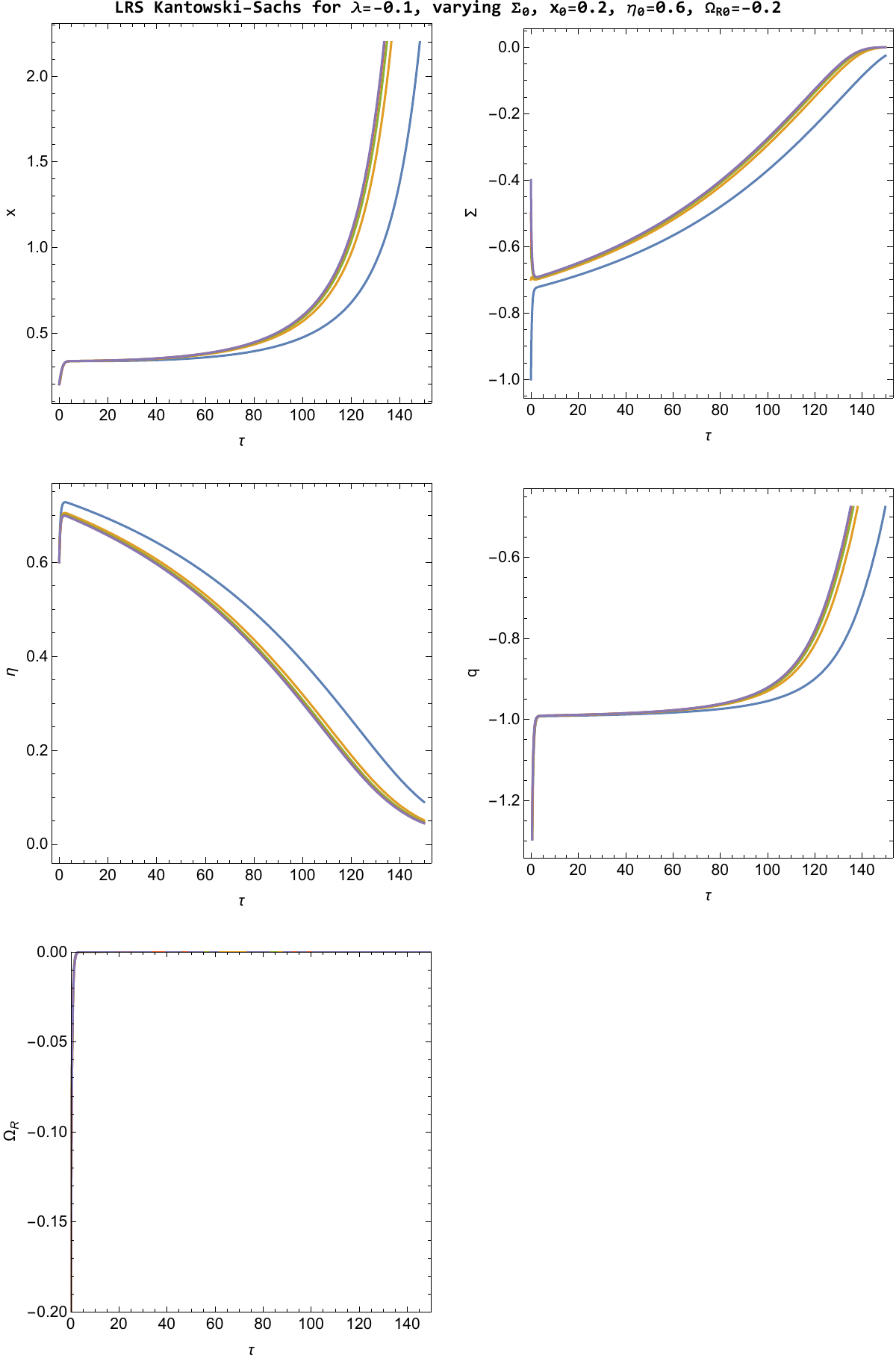}
\caption{Kantowski-Sachs: Numerical solution of the field equations within the Kantowski-Sachs background for different set of initial conditions. For the numerical simulations we selected $\lambda=-0.1$. For the initial conditions we consider $x_{0}=0.2$ , $\Sigma_{0} = \left\{ -1,-0.7,-0.6,-0.5.-0.4 \right\} $, $\eta_{0}=0.6$ and $\Omega_{\kappa 0}=-0.2$. } %
\label{ksp2}%
\end{figure}

In the finite regime no attractors were observed which describe other physical solutions and so as before the model unfortunately suffers from Big Rip and Big Crunch singularities.

\section{Conclusion} \label{sec4}

In this work, we explored the evolution of cosmological anisotropic spacetimes in four-dimensional $f\left( G\right) $-gravity. We show that this second-order gravitational theory is equivalent to Einstein-Gauss-Bonnet scalar field gravity without the kinetic term, where the scalar field carries the geometrodynamical degrees of freedom provided by the nonlinear function $f\left( G\right)$.

We made use of the Misner variables and considered a generic line element with two scale factors, which admitted four isometries. The Lie algebra formed by these isometries depends on the sign of a free parameter. The possible Lie algebras are $A_{1}\otimes A_{3,4}$, $A_{1}\otimes A_{3,5}$, and $A_{4,10}$ in the Patera et al. classification scheme \cite{patera}. Consequently, this geometric line element describes the LRS Bianchi I, LRS Bianchi III, and Kantowski-Sachs spacetimes. These geometries have the property of admitting a minisuperspace description. Indeed, we calculated the Ricci and Gauss-Bonnet scalars in terms of the metric functions and by substituting them into the gravitational Action Integral and integrating by parts to eliminate boundary terms, obtained a point-like Lagrangian.

In order to understand the effects of the modified Gauss-Bonnet theory on the evolution of anisotropies, we employ dimensionless variables and numerically explore the phase space. We consider a specific function $f\left( G\right) $ such that the corresponding scalar field potential is an exponential function. In this simple case, the maximum dimension of the resulting dynamical system is four, for the Bianchi III and Kantowski-Sachs geometries, and three for the Bianchi I geometry.

From the numerical solutions, we conclude that the field equations in the finite regime, regardless of the spatial curvature of the two-dimensional space, admit as attractors a point that describes an isotropic and spatially flat FLRW scaling solution, as well as the Minkowski spacetime. On the other hand, Big Rip or Big Crunch singularities appear in the infinite regime.

This analysis shows that $f\left( G\right) $-gravity can address the flatness and isotropy problems, and that there exist trajectories capable of describing two inflationary eras: a de Sitter epoch and a scaling acceleration solution. However, it is important to mention that Kasner and Kasner-like solutions are unstable. Moreover, the existence of the Minkowski solution as
an attractor indicates the possible presence of naked singularities in the Kantowski-Sachs geometry.

In future work, we plan to extend this study to the case of static spherically symmetric geometries and to investigate the existence and stability properties of black hole solutions. \\

\textbf{Data Availability Statements:} Data sharing is not applicable to this article as no datasets were generated or analyzed during the current study. \\
\\
\textbf{Code Availability Statements:} Code sharing is available after request.
\\
\begin{acknowledgments}
AG was supported by Proyecto Fondecyt Regular 1240247. GL \& AP were supported by Proyecto Fondecyt Regular 2024, Folio 1240514, Etapa 2025.
\end{acknowledgments}

\end{document}